# Field-induced high coercive ferromagnetic state and magnetoresistance in the antiferromagnetically ordered compound $Fe_{0.5}TiS_2$

N V Baranov[1,2], E M Sherokalova[2], A S Volegov[2], AV Proshkin[1,2], N V Selezneva[2] and E P Proskurina[2],

[1] Institute of Metal Physics, Russian Academy of Science, 620990, Ekaterinburg, Russia;

[2] Institute of Natural Science, Ural Federal University, 620083, Ekaterinburg, Russia

**Abstract.** The measurements of the magnetic susceptibility, magnetization, electrical resistivity and magnetoresistance have been performed for the Fe intercalated compound $Fe_{0.5}TiS_2$. According to X-ray diffraction measurements the $Fe_{0.5}TiS_2$ compound synthesized in the present work has a monoclinic crystal structure (space group $I12/m1$) which results from the ordering of Fe ions and vacancies between S-Ti-S tri-layres. The changes in the heat-treatment conditions at temperatures below 1100 °C do not lead to an order-disorder transition within the subsystem of intercalated Fe ions. It has been shown that this compound exhibits an antiferromagnetic (AF) ground state below the Neel temperature $T_N \approx 140$ K. Application of the magnetic field at $T < T_N$ induces a metamagnetic phase transition to the ferromagnetic (F) state, which is accompanied by the large magnetoresistance effect ($|\Delta\rho/\rho|$ up to 27 %). The field-induced AF-F transition is found to be irreversible below ~ 100 K. The magnetization reversal in the metastable F state at low temperatures is accompanied by substantial hysteresis ($\Delta H \sim 100$ kOe) which is associated with the Ising character of Fe ions.

## 1. Introduction

After intercalation, the transition metal (T) dichalcogenides $TX_2$ (X = S, Se, Te) with a layer crystal structure may exhibit substantially different physical properties in comparison with the parent compounds [1, 2]. For instance, the intercalation of $TiSe_2$ with Cu leads to the appearance of the superconductivty and suppresses the charge density wave [3], while after Fe- or Cr-intercalation the $Fe_{0.5}TiSe_2$ [4, 5] and $Cr_{0.5}TiSe_2$ [6,7] compounds reveal antiferromagnetic (AF) properties, and a highly coercive ferromagnetic (F) behavior is observed in $Fe_{0.25}TaS_2$ [8]. The intercalation of guest (M) atoms or molecules into $TX_2$ matrixes is possible owing to the presence of a weak coupling between X-T-X tri-layers via van der Waals (vdW) forces. The intercalated M atoms usually occupy the octahedrally coordinated sites in vdW gaps of $M_xTX_2$ compounds. When the M atoms have a magnetic moment the $M_xTX_2$ compounds can be considered as an analog of artificial multi-layer structures since the M layers are separated by non-magnetic layers. The long-range magnetic order was observed in some $M_xTX_2$ compounds at $x \geq 0.25$, while the compounds with lower M concentrations exhibit spin-glass or cluster-glass magnetic states [4-12]. The magnetic order in highly intercalated $M_xTX_2$ systems depends on both the type of M atoms and $TX_2$ matrix. Thus, the cromium intercalated compounds $Cr_{0.33}TX_2$ based on niobium diselenide and disulfide show different magnetic orders: a long range ferromagnetic order is observed in $Cr_{0.33}NbSe_2$ [10], whereas $Cr_{0.33}NbS_2$ exhibits a long period helical spin structure [11]. The appearance of a three-dimensional magnetic order in highly intercalated $M_xTiX_2$ compounds is accounted for a combination of two types of indirect exchange interactions: i) the indirect exchange interaction between 3d electrons of M atoms via conduction electrons, and ii) the superexchange interaction through chalcogen atoms [12-14]. The first exchange interaction of the Ruderman-Kittel-Kasuya-Yoshida (RKKY) type dominates apparently within the layers of the inserted M atoms since the $M_xTX_2$ compounds exhibit a substantially higher conductivity parallel to the layers in comparison with conductivity in the perpendicular direction [1]. The preferential orientation of magnetic moments of intercalated 3d atoms and magnetic anisotropy of $M_xTX_2$ compounds is observed to depend on the type of M atoms inserted into $TX_2$ matrix [12]. The presence of a non-zero orbital moment in 3d metal ions intercalated into $TX_2$ matrixes may lead to the substantial magnetocrystalline anisotropy and high coercivty as revealed in some $M_xTX_2$ compounds by magnetization measurements on single crystalline samples. Thus, in the earlier work of Eibschütz et al [8], the value of the anisotropy field for the ferromagnetic compounds $Fe_{0.28}TaS_2$ was found to be about 500 kOe and the coercive field $H_c \approx 55$ kOe. These data were confirmed recently by magnetization measurements on the single crystalline $Fe_{0.25}TaS_2$ sample with close Fe concentration [15]. The coercive field of the same order (~50 kOe) was observed for $Fe_{0.5}TiS_2$ [16]. In both $Fe_{0.5}TiS_2$ and $Fe_{0.25}TaS_2$, the high coercive force can be explained by the presence of non-zero orbital moment of $Fe^{2+}$ ions, as revealed by X-Ray MCD measurements [17, 18]. As to the magnetic state of $Fe_{0.5}TiS_2$, there are contradictory data in the literature. The magnetic state of $Fe_{0.5}TiS_2$ at low temperatures was proposed to be antiferromagnetic with inclusion ferromagnetic clusters [19] antiferromagnetic [20], or ferromagnetic [16, 21, 22]. The phase diagrams presented in [9, 21, 22] for the $Fe_xTiS_2$ imply the presence of long-range ferromagnetic order at the Fe concentration above $x \approx 0.4$, however, some indications for the presence of a more complicated magnetic structure in $Fe_{0.5}TiS_2$ have been revealed by preliminary neutron diffraction experiments [23]. The magnetic ordering in $Fe_{0.5}TiS_2$ was suggested to depend on the order degree of Fe ions and vacancies within vdW gaps and deviations from the stochiometry [16, 20, 24]. At low intercalant concentrations, the magnetic state of $Fe_xTiS_2$ is classified as an Ising spin-glass ($x \leq 0.20$) or cluster-glass ($0.20 < x < 0.40$) [9, 25].

The main purpose of the present work is to determine the ground magnetic state of the intercalated compound $Fe_{0.5}TiS_2$. In order to answer the question what type of the magnetic order exists in $Fe_{0.5}TiS_2$ together with detailed magnetization measurements we have measured the field- and temperature dependences of the electrical resistivity suggesting that the conduction electron scattering will be sensitive to the mutual orientation and periodicity in the arrangement of Fe magnetic moments owing to the s–d exchange interaction. Another goal of the paper is to achieve a better understanding of the origins of the high coercivity in the compounds the magnetic properties of which are determined by Fe ions.

2. **Experimental details**

The $Fe_{0.5}TiS_2$ compound was prepared by chemical reaction inside a sealed quartz tubes in two stages. At first, the parent compounds $TiS_2$ was synthesized by heat treatment of a mixture of starting materials at 800 ºC during one week. At the second stage the mixtures of Fe and $TiS_2$-powders were pressed into cylinder pellets and annealed at the same conditions. The obtained specimen were milled, compacted into tablets and then homogenized for two weeks at 800 °C followed by cooling down through removal of the tube from the oven into air. In order to check the suggestion [26] about the presence of the order-disorder transition in $Fe_{0.5}TiS_2$ we have additionally performed heat treatments at 350 ºC and 1100 ºC.

The as-synthesized and heat-treated samples were examined by powder X-ray diffraction analysis by using a Bruker D8 Advance diffractometer with Cu $K_\alpha$ radiation. For Rietveld refinements of the crystal structure, the FULLPROF program was used [27]. The magnetization measurements were performed on the pressed samples (4 mm diameter and 1 mm thickness) by means of a Quantum Design SQUID magnetometer in the temperature interval from 2 up to 300 K. The magnetic field was directed perpendicular to the sample plane.

The temperature and field dependences of the electrical resistivity of $Fe_{0.5}TiS_2$ were measured from 4 K up to 300 K by a four-contact ac method by using a cryo-free DMS-1000 system (Dryogenic Ltd, UK) in magnetic fields up to12 T.

## 3. Results

### 3.1. Crystal structure

The crystal structure of the as-synthesized $Fe_{0.5}TiS_2$ sample was identified as of the monoclinic $Cr_3S_4$-type

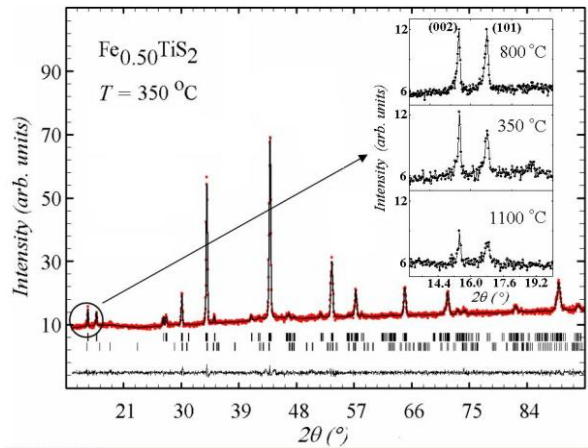

Figure 1. Observed at room temperature (symbols) and calculated (line) X-ray diffraction pattern for the $Fe_{0.5}TiS_2$ sample after additional two-weak heat treatment at 350 °C. The positions of the Bragg peaks are marked at the bottom for the main $Fe_{0.5}TiS_2$ monoclinic phase (upper row) as well as for the FeS compound as for a foreign phase (lower row). The difference between calculated and observed intensities is shown in the bottom of the figure. Insets show the (002) and (101) reflections for the as-synthesized (800 °C), heat-treated at 350 °C and 1100°C samples.

(space group $I12/m1$) with the lattice parameters: $a = a_0\sqrt{3} = 5.950$ Å, $b = a_0 = 3.421$ Å, $c = 2c_0 = 11.497$ Å, $\beta = 89.87°$, where $a_0$ and $c_0$ are the hexagonal cell parameters. These results are in agreement with published data [28]. We did not observe any extra reflections on the x-ray diffraction pattern for this sample. The structure of $Fe_{0.5}TiS_2$ can be considered as a superstructure of the parent compound $TiS_2$ (space group $P\bar{3}m1$) which results from the ordering of Fe atoms and vacancies within layers located between the three-layer S-Ti-S blocks with hexagonal packing of the sulfur and titanium atoms. As was suggested in [26, 28], the order degree in the $Fe_{0.5}TiS_2$ compound can be estimated by using the ratio $I_{(002)}/I_{(101)}$ of the intensities of (002) and (101) reflections. As follows from the inset in figure 1, these two reflections have nearly the same intensity for the as-synthesized at 800 °C, which is indicative of a high order degree of Fe atoms and vacancies in our sample. In figure 1, we plotted the X-ray diffraction pattern for the $Fe_{0.5}TiS_2$ sample which was additionally heat-treated at 350 °C during two weeks. The low-temperature heat treatment of the sample is observed to lead to the appearance of additional reflections of low intensities. These extra reflections do not relate to the $Cr_3S_4$-type structure. Our analysis has shown that some amount (~ 3%) of the FeS compound as a foreign phase may exist in our sample ($P\bar{6}2c$ space group, lattice parameters: $a = 5.923$ Å and $c = 11.568$ Å). It is evidently that the appearance of a new phase after heat treatment at 350 °C should result in some deviation from the stoichiometry of the $Fe_{0.5}TiS_2$ phase as well to the change of the $I_{(002)}/I_{(101)}$ intensity ratio. The sample heat treated 350 °C at exhibits an increased $I_{(002)}/I_{(101)}$ ratio (see inset in figure 1) and the following lattice parameters: $a = 5.945$ Å; $b = 3.417$ Å; $c = 11.495$ Å and $\beta = 90.31°$. In order to check the assumption [26] about the presence of the order-disorder transition within subsystem of Fe ions and vacancies in $Fe_{0.5}TiS_2$ with increasing temperature above 450 °C we have also performed heat treatment at temperature 1100 °C during 3 hours with subsequent rapid cooling to room temperature by removal of the tube with the sample from the oven into air. After such a procedure, we did not observe the change of the crystal structure of our compound from the monoclinic $Cr_3S_4$-type to the hexagonal $CdI_2$-type crystal structure (space group $P\bar{3}m1$) with statistical distribution of Fe atoms. As can be seen in the inset in figure 1, the (002) and (101) reflections are still observed for $Fe_{0.5}TiS_2$ after heat treatment at 1100 °C, while in the case of $CdI_2$-type crystal structure, the peaks with such indexes should be absent in this angle interval. Our data are in agreement with results [28] which show that any changes in the heat-treatment and cooling conditions of $Fe_{0.5}TiS_2$ ($FeTi_2S_4$) below 1300°C do not lead to complete disordering of vacancies and Fe atoms.

*3.2. Magnetic susceptibility and magnetization measurements*

Figures 2(a) and 2(b) show the temperature dependences of the magnetization measured on a polycrystalline as-synthesized $Fe_{0.5}TiS_2$ sample at different applied magnetic fields and cooling/warming regimes. The measurements at $H = 1$ kOe have revealed a pronounced peak of the magnetization at $T \approx$ 140 K on the ZFC curve (figure 2(a)). During cooling in the same field, some difference in the $M(T)$ behavior was observed below this temperature. This difference becomes more appreciable at the measurements in higher fields (see figure 2(b)). The peak position shifts towards lower temperature with increasing field. Thus, at $H = 10$ kOe, the maximum of the magnetization measured at the ZFC regime is observed at $T = 124$ K. Whereas the $M(T)$ dependence for $Fe_{0.5}TiS_2$ obtained at low fields ($H \sim 1$ kOe or less) at the ZFC regime may be deemed to be quite typical for AF ordered compounds, the change of the magnetization with temperature at $H = 70$ kOe can be associated with the presence of a ferromagnetic order as well. Our value of the magnetic critical temperature 140 K for this compound is in agreement with the Curie or Neel temperatures reported in the literature ($T_N = 132$ K [19], $T_N = 138$ K [20], $T_C = 111$ K [29], $T_C = 125$ K [16]. As is shown in the inset in figure 2(a), the magnetic susceptibility of $Fe_{0.5}TiS_2$ obeys the Curie-Weiss law $\chi(T) = \chi_0 + C/(T - \Theta_p)$ at temperatures above 220 K ($\chi_0$ is the temperature-independent term). The effective magnetic moment per Fe atom is estimated to be 3.62 $\mu_B$ which is lower than $\mu_{eff} = 4.89\ \mu_B$ expected for a high-spin state of the $Fe^{2+}$ ion with $g = 2$. Our value is in agreement with the data obtained on a single crystalline sample (3.95 $\mu_B$ and 3.45 $\mu_B$ for the directions parallel and perpendicular to the *c*- axis, respectively [20]). The reduced $\mu_{eff}$ value may be associated with the hybridization of 3d electrons of intercalated Fe ions with Ti 3d states and S 3p states in $TiS_2$. The paramagnetic Curie temperature $\Theta_p$ is found to be equal about of 123 K, which indicates that the ferromagnetic exchange interactions dominate in $Fe_{0.5}TiS_2$. The positive $\Theta_p$ values were also observed for other $Fe_xTiS_2$ compounds with lower Fe concentrations ($0 < x \leq 0.42$) [30].

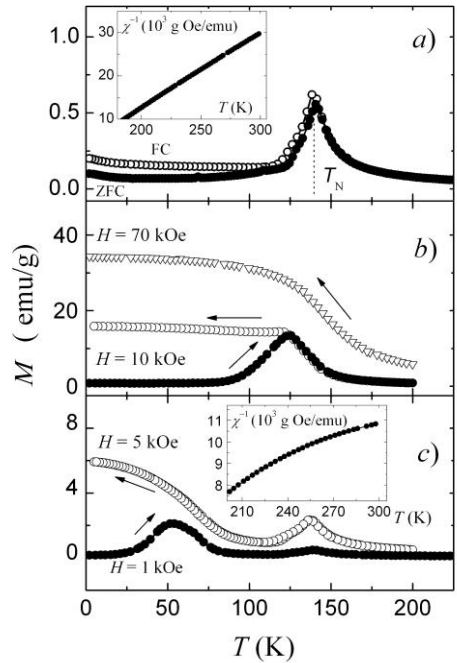

Figure 2. Temperature dependences of the magnetization for the as-synthesized (*a,b*) and for the heat-treated at 350 °C (*c*) samples of $Fe_{0.5}TiS_2$ at various fields and cooling/warming conditions. Insets show the temperature dependence of the reciprocal susceptibility.

In figure 2(c) we plotted the temperature dependences of the magnetization measured on the $Fe_{0.5}TiS_2$ sample after additional two-weak heat treatment at 350 °C. As it turned out, this heat-treated sample exhibits a substantially different temperature behavior of the magnetization in comparison with the non-heat-treated sample. Below 75 K, the heat-treated sample shows a pronounced magnetization hump and the difference between ZFC and FC magnetization. Such kind of the magnetization behavior in $Fe_{0.5}TiS_2$ was observed in earlier articles (see Ref. [19], for instance). The magnetization hump on the ZFC curve below 75 K (figure 2(c)) can hardly be associated with the $Fe_{1-x}S$ phase which may appear in our sample after 350 °C heat treatment since FeS shows an antiferromagnetic behavior below ~ 600 K [31] and does not exhibit any magnetic transformations at $T < 100$ K. The above mentioned deviation from the stoichiometry may cause the absence of full compensation of the magnetization in neighbor Fe layers and frustrations in the Fe-Fe exchange interactions. Therefore, the maximum of the magnetization below 75 K observed in $Fe_{0.5}TiS_2$ after low-temperature heat treatment may be associated the appearance of magnetic clusters with short-range ferromagnetic order within the AF ordered matrix. It should be noted that the magnetic susceptibility of the $Fe_{0.5}TiSe_2$ sample heat-treated at 350 °C does not follow the Curie-Weiss law (shown in the inset in figure 2(c)) unlike the as-synthesized compound. Such a behavior may result from the presence of short-range magnetic correlations above $T_N$ and a small amount of the AF ordered FeS phase.

In figure 3, we collected the *M*(*H*) dependences measured at various temperatures on the as-synthesized $Fe_{0.5}TiSe_2$ sample. At $T = 130$ K, i.e. just below the Neel temperature, application of a magnetic field above the critical value $H_c^{up}$ ~ 5 kOe leads to the sharp growth of the magnetization till reaching the saturation (figure 3(a)). Such a behavior may be attributed to the field-induced AF-F transition of the first order within the subsystem of Fe magnetic moments. This metamagnetic transition is accompanied by hysteresis which increases with decreasing temperature. In order to avoid the prehistory effect the measurements at each temperature were performed after heating the sample up to 150 K with following cooling down at zero field. Figure 4 shows the temperature dependences of the critical transition fields $H_c^{up}$ and $H_c^{down}$ when the field increases and decreases, respectively. As is seen, when the temperature decreases below ~100 K the width of the hysteresis loop $\Delta H = H_c^{up} - H_c^{down}$ exceeds the critical field $H_c^{up}$ and the AF-F transition becomes irreversible. Below 100 K, $H_c^{down}$ was taken to be equal to the coercive field, $H_c$. The $\Delta H$ value (shown in the inset in figure 4) decreases with increasing temperature from ~100 kOe at low temperatures to nearly zero at $T$ ~ 130 K apparently because of the thermally activated motion of domain walls.

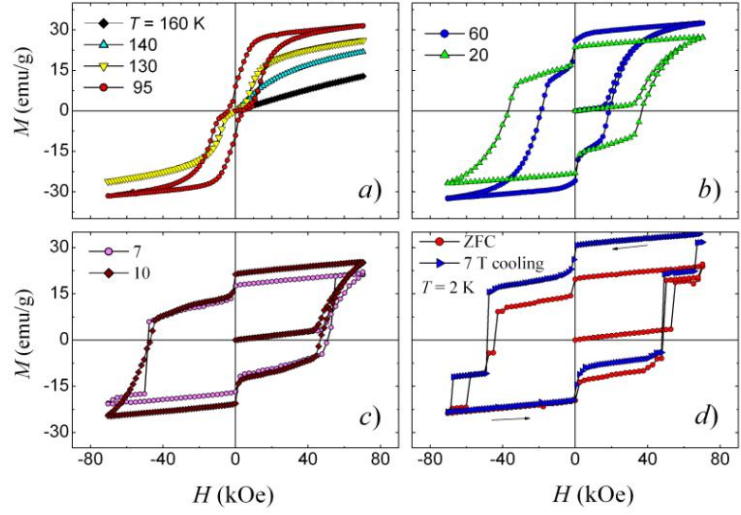

Figure 3. Field dependences of the magnetization for the as-synthesized $Fe_{0.5}TiS_2$ sample measured at various temperatures.

The *M*(*H*) isotherms measured above 7 K display the smooth behavior across the critical transition field while below this temperature, the magnetization jumps are observed around the coercive field (see figure 3(c) and figure 3(d)). It should be noted that such a crossover from the gradual change of the magnetization to a step-like behavior with decreasing temperature below ~5−10 K was observed in different magnetic systems, in particular, at the metamagnetic phase transitions in the substituted intermetallic compound CeFe$_2$ [32], in manganites [33] as well as in amorphous alloys [34]. When the sample was cooled down to 2 K in an applied field of 7 T, the substantially higher magnetization value was obtained at this temperature in comparison with the ZFC case (see figure 3(d)). However, this difference disappears when the 7 T field is applied at *T* = 2 K in opposite direction. In general, our results on the magnetization behavior of Fe$_{0.5}$TiS$_2$ are in agreement with those obtained previously in [16], where the ground magnetic state of this compound was characterized as ferromagnetic.

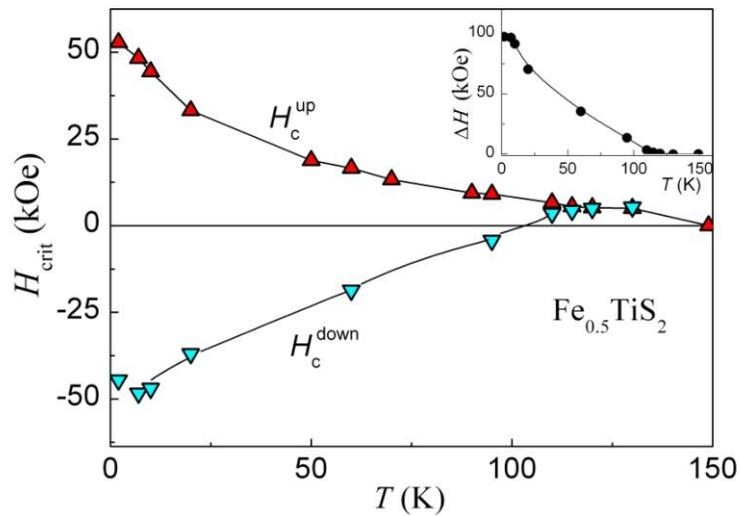

Figure 4. Temperature dependences of the critical transition fields $H_c^{up}$ and $H_c^{down}$ for the as-synthesized Fe$_{0.5}$TiS$_2$ sample when the field increases and decreases, respectively. Inset shows the temperature variation of the width of the hysteresis loop.

As to the Fe$_{0.5}$TiS$_2$ sample heat-treated at 350 °C, only small differences in the magnetization behavior were observed in comparison with the as-synthesized sample (figure 5). The emergence of a small ferromagnetic-like contribution on the initial magnetization curve was detected after cooling below 75 K (marked by circles). The measurements of the magnetization at *T* =10 K in the field interval ± 20 kOe have revealed a partial hysteresis loop (figure 5(c)) which can be associated with the magnetization reversal in small regions (clusters) with short-range Fe-Fe ferromagnetic correlations. At temperatures below 10 K, staircase-like $M(H)$ dependences are observed for the heat-treated sample as well.

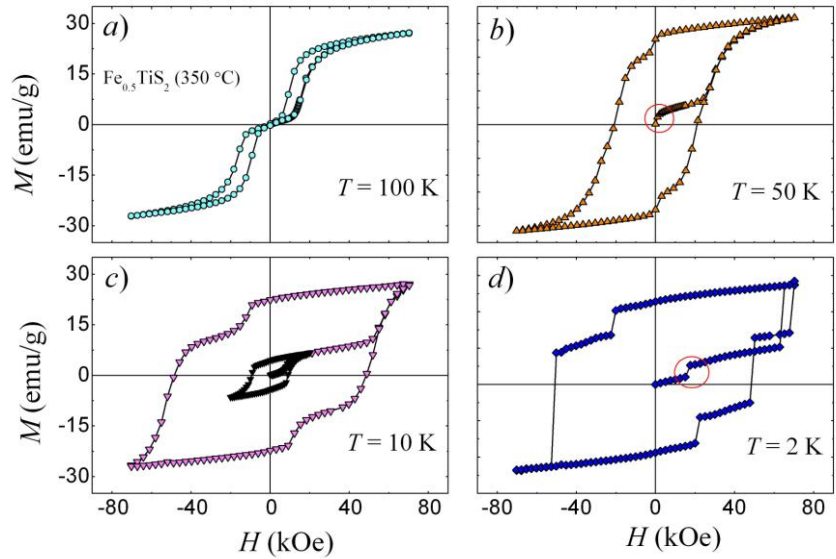

Figure 5. The $M$ versus $H$ isotherms measured on the $Fe_{0.5}TiSe_2$ sample heat-treated at 350 °C. The steps on the $M(H)$ curves associated with the magnetization process within clusters are marked by circles. The partial hysteresis loop is shown by full symbols.

*3.3. Electrical resistivity and magnetoresistance*

As can be seen from figure 5(a), the electrical resistivity of the as-synthesized sample $Fe_{0.5}TiSe_2$ exhibits a metallic behavior in the magnetically ordered state while in the paramagnetic region, the resistivity value remains nearly constant with increasing temperature from 150 K up to 300 K, which agrees with previously reported data [35]. The pronounced change in the $\rho(T)$ dependence around the magnetic ordering temperature indicates a strong interaction of conduction electrons with magnetic moments of iron atoms. The behavior of the resistivity of $Fe_{0.5}TiSe_2$ differs from that observed for the parent compound $TiS_2$, which shows a monotonous growth of the resistivity with increasing temperature [2]. In the case of $Fe_{0.5}TiS_2$, the resistivity behavior above $T_N$ may be influenced by short-range magnetic correlations which persist apparently in the paramagnetic region. This suggestion is supported by the fact that the magnetic susceptibility does not strictly follow the Curie-Weiss up to 220 K

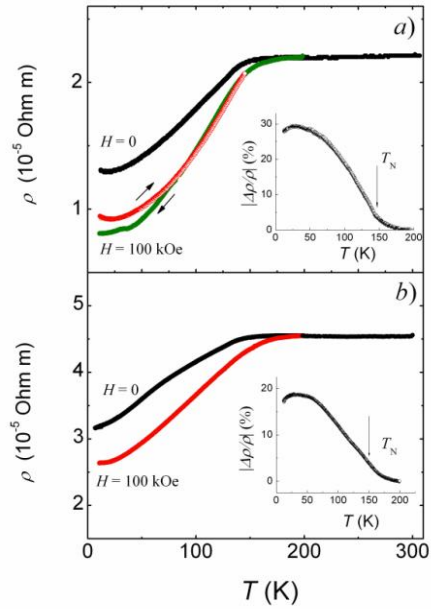

Figure 6. Temperature dependences of the electrical resistivity for the as synthesized (*a*) and heat-treated at 350 °C (*b*) $Fe_{0.5}TiS_2$ samples measured at $H = 0$ and at $H = 100$ kOe with increasing and decreasing temperature. Inset shows the temperature variation of the magnetoresistance measured at $H = 100$ kOe.

(see inset in figure 2(a)). The $Fe_{0.5}TiS_2$ sample heat-treated at 350 °C shows higher resistivity values apparently because of the presence of a foreign phase, although the $\rho(T)$ dependence (figure 6(b)) shows nearly the same shape as for the as-synthesized sample. Some difference may be seen in the temperature interval 40 K $< T <$ 100 K where the $\rho(T)$ dependence for the heat-treated sample demonstrates the negative curvature unlike positive curvature $\rho$ versus $T$ dependence for the as-synthesized sample in the same temperature range. Such a difference may result from the presence of an additional contribution to the resistivity associated with appearance of short-range magnetic correlations in magnetic clusters in the heat-treated sample when the temperature decreases below 75 K. The presence of the contribution of such a kind was revealed in some diluted alloys with magnetic clustering [36]. For both samples we have also measured the $\rho(T)$ dependences under application of the 100 kOe magnetic field with increasing (ZFC regime) and decreasing temperature (FC regime). As it turned out, the application of a high magnetic field substantially reduces the electrical resistivity below the magnetic ordering temperature. As follows from figure 6(a) and 6(b), the decrease of the resistivity in comparison with $\rho$ value at $H = 0$ is more pronounced when the measurements were made at the FC regime (figure 6(a)). The lower value of the resistivity observed at low temperatures in the FC case is consistent with a higher magnetization obtained after cooling of the $Fe_{0.5}TiS_2$ sample at $H = 70$ kOe (see figure 3(d)). As follows from the insets in figure 6, the absolute value of the magnetoresistance $|\Delta\rho/\rho| = |[\rho(0) - \rho(H)]/\rho(0)|$ of the as synthesized sample increases with lowering temperature and reaches ~ 30% at low temperatures. The reduced $|\Delta\rho/\rho|$ values obtained for the heat treated-sample (inset in figure 6(b)) in comparison with the as-synthesized sample is associated with increased $\rho(0)$. It should be emphasized here, that the temperature dependences of $\Delta\rho/\rho$ for both samples are in contradiction with the magnetoresistance behavior observed in ferromagnetic materials which usually demonstrate the maximal absolute value of $\Delta\rho/\rho$ in the vicinity of the Curie temperature. The giant magnetoresistance (GMR) is often observed in metallic antiferromagnets based on rare earth [37, 38] and 3d transition metals [39, 40] while the bulk ferromagnetically ordered materials as a rule do not exhibit significant changes in the electrical resistivity well below $T_C$.

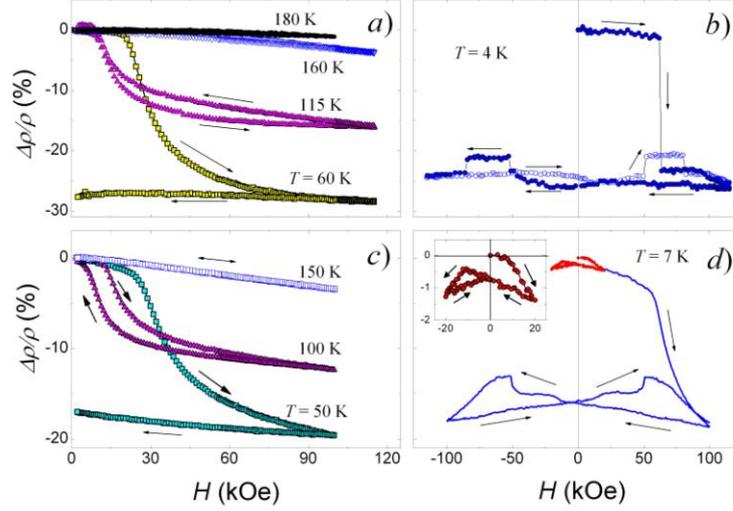

Figure 7. The magnetoresistance isotherms for the as-synthesized (*a*, *b*) and heat-treated (*c*, *d*) samples of $Fe_{0.5}TiS_2$. Insert shows the partial hysteresis loop measured on the heat-treated sample in detail.

The field dependences of the magnetoresistance of both samples measured at various temperatures are shown in figure 7. As is seen, the changes in the resistivity under application of the magnetic field are qualitatively consistent with the magnetization behavior. Just below the magnetic ordering temperature, an increase of the magnetic field above a critical value leads to an accelerated resistivity decrease which accompanies the metamagnetic phase transition (figures 7(a) and 7(c)). The hysteretic behavior in the vicinity of the critical field becomes more pronounced with lowering temperature. The irreversibility in the change of the resistivity is clear seen in figure 7(b) and 7(d) which display the $\Delta\rho/\rho$ versus $H$ dependences for both the as-synthesized and heat-treated sample, respectively. For as-synthesized sample measured at $T = 4$ K, a sharp drop in the resistivity is observed when the field reaches a critical value while further variation in the field within ±120 kOe leads to a hysteresis loop in which the $\Delta\rho/\rho$ value varies between -27% and -21% (figure 7(b)). The step-like changes of the resistivity take place at the same fields at which the magnetization jumps are observed. As follows from figure 7(b), the $\Delta\rho(H)/\rho$ dependence exhibits a pronounced stepwise behavior in the vicinity of the coercive field. At low temperatures, switching off the field does not restore the initial resistivity values of both samples. The samples can be returned to its initial state only after heating to $T > 110$ K and subsequent cooling in zero field. Generally, the resistivity behavior of the heat-treated sample under application of the magnetic field is analogous to that observed for the as-synthesized sample, which allows us to conclude that the heat treatment does not affect substantially the ground magnetic state of the compound. Some difference in the $\Delta\rho(H)/\rho$ dependence of heat-treated sample is revealed at low temperatures (figure 7(d)). The magnetoresistance shows the partial hysteresis loop when the applied field varies within field interval ± 20 kOe, which may be associated with the magnetization reversal within clusters. The presence of such loop is in agreement with the magnetization change presented in figure 5(c).

## 4. Discussion

The analysis of the crystal structure of $Fe_{0.5}TiS_2$ samples studied in the present work shows that the Fe ions and vacancies remain always in the ordered state within the layers between S-Ti-S sandwiches despite different heat treatments at temperatures below 1100 °C. The low-temperature (below 500°C) heat treatment of $Fe_{0.5}TiS_2$ samples may lead to the appearance of a second phase (of the FeS type) and to the deviation from the stoichiometry of the main phase. Only slight decrease of the order degree in the $FeTi_2S_4$ sample cooled from 1100°C was observed (see inset in figure 1) in comparison with as-synthesized sample, which is consistent with the data presented in [24]. The order-disorder transition in the vicinity of 450 °C proposed by Muranaka [26] seems to be indeed absent in $Fe_{0.5}TiS_2$. Bearing in mind our heat-treatment experiments the changes in structural and magnetic characteristics reported in [26] can be rather ascribed to the phase transformations in $Fe_{0.5}TiS_2$ samples at the low-temperature heat treatments (below 500 K) and to the stoichiometry deviations in the ordered $Fe_{0.5}TiS_2$ phase.

The previous contradictory data [19-22] about the magnetic state of $Fe_{0.5}TiS_2$ as well as the present study show how difficult to identify the type of a magnetic order which is realized in this compound at low temperatures by using the results of the magnetization measurements only. This is because the $Fe_{0.5}TiS_2$ compound being magnetized in high magnetic fields behaves at low temperatures as a high-coercive ferromagnetic material. However, the GMR effect which is observed in the whole temperature range below $T_N$ for both as-synthesized and heat-treated samples we consider as a clear evidence of the antiferromagnetic ground state in $Fe_{0.5}TiS_2$. More appreciable resistivity change at the AF-F transition is expected in the case of single crystalline samples. The GMR at the field-induced AF-F transitions in AF ordered compounds may originates in the reconstruction of the Fermi surface due to the disappearance of energy gaps on superzone boundaries [41]. The presence of energy gaps in the electron spectrum and the increased electrical resistivity in the AF ordered compounds with metallic-type conductivity originates in a larger period of the antiferromagnetic structure in comparison with the period of the crystal lattice. Therefore, one may suggest that the magnetic unit cell of our $Fe_{0.5}TiS_2$ samples is larger than the crystal unit cell.

From the susceptibility measurements performed in earlier studies on single crystalline samples of $Fe_xTiS_2$ compounds ($x<0.35$) with disordered Fe atoms, it was concluded that Fe magnetic moments lie along the $c$ axis [25]. Such an orientation may be associated with an influence of the crystal electric field and spin-orbital interactions since Fe ions in $Fe_xTiS_2$ are observed to exhibit a non-zero orbital moment [17]. As shown by X-ray MCD experiments performed for $Fe_xTiS_2$ ($0 < x \leq 0.33$), the Fe ions in $Fe_{0.33}TiS_2$ exhibit an orbital magnetic moment of about 0.3 $\mu_B$ [17]. It is worth to note that unlike Fe moments in $Fe_xTX_2$ the magnetic moments of $Mn^{2+}$ ions in $Mn_xTX_2$ (T = Nb, Ta) are observed to lie parallel to the layers [42]. Such a difference may be explained by the fact that $Mn^{2+}$ ions in these compounds are not influenced by the crystal electric field because of spherical symmetry of the $3d^5$ electron density. Some deviation of the magnetic moment direction from the $c$-axis (~ 15º) is observed in the antiferromagnetic compound $Fe_{0.5}TiSe_2$ [5], which was ascribed to the distortion of the tetrahedron formed by nearest Se ions surrounding the $Fe^{2+}$ ion. These distortions are attributed to the monoclinicity of the crystal structure of $Fe_{0.5}TiSe_2$ caused by formation of the one-dimensional Fe chains along the $b$ axis. Bearing in mind that our $Fe_{0.5}TiS_2$ sample exhibits an analogous crystal structure some deviations of Fe magnetic moments from the *c* axis may be expected in this compound as well. Another suggestion in respect to the peculiarities of the AF order in $Fe_{0.5}TiS_2$ below $T_N$ can be made when analyzing the magnetization and magnetoresistance data obtained in the present work. Despite both the Fe intercalated compounds based on the titanium diselinide and titanium disulfide have an antiferromagnetic alignment of Fe magnetic moments and nearly the same magnetic ordering temperatures ($T_N$ = 140 K for $Fe_{0.5}TiS_2$ (see above) and $T_N$ = 135 K for $Fe_{0.5}TiSe_2$ [5]) one can suggest that their magnetic structures are substantially different. In the $Fe_{0.5}TiSe_2$ compound, the magnetic moments of iron atoms located in the same layer are coupled antiferromagnetically with each other [5]. Fairly high magnetic fields are needed to overcome this strong AF exchange interaction within Fe layers and to destroy the AF order. Indeed, the AF-F transition in $Fe_{0.5}TiSe_2$ was observed to start at a critical field of about 300 kOe at $T$ = 4.2 K [5]. This critical transition fields is several times larger than that observed in $Fe_{0.5}TiS_2$, which allows us to suggest that unlike the selenium based counterpart the $Fe_{0.5}TiS_2$ compound exhibits a layered magnetic structure with a weak antiferromagnetic exchange field ($H_{ex1}$) between neighbor Fe layers and much stronger Fe-Fe ferromagnetic exchange ($H_{ex2}$) within layers ($|H_{ex2}| \gg |H_{ex1}|$). The field induced AF-F phase transformations and large hysteresis observed in $Fe_{0.5}TiS_2$ originates apparently in such a layered antiferromagnetic structure together with the Ising character of Fe ions. As is known [43], the metamagnetic phase transitions under application of a magnetic field in a simple two-sublattice antiferromagnet take place in the case when the anisotropy field, $H_a$, is higher than the intersublattice exchange field $H_{ex1}$. Application of a magnetic field just below $T_N$ leads to the AF-F transition of second order while this transition becomes of the first-order type with lowering temperature below tri-critical point [43]. In the last case, the AF-F transition occurs through the nucleation of the F phase within the AF matrix followed by motion of the interphase boundaries. When the anisotropy field in an antiferromagnet surpasses not only the intersublattice (interlayer) exchange but also the intrasublattice (intralayer) exchange interaction ($H_a > H_{ex2} \gg H_{ex1}$) the propagation field which is needed for the motion of the AF-F interphase boundary will be controlled by the total local exchange field acting on Fe moments located at the boundary and by thermal activation. In such case, when the temperature decreases to zero, the limit of the coercive field will be determined mainly by intralayer ferromagnetic exchange field $H_{ex2}$. However, the growth of the coercive field with decreasing temperature below a few Kelvin may be ceased due to the crossover from the thermal activation of the domain wall motion to macroscopic quantum tunneling (MQT) [44]. In $Fe_{0.5}TiS_2$, the change in the magnetization and magnetoresistance behaviors with decreasing temperature below 10 K in the vicinity of the coercive field (see figures 3, 5 and 7) may be associated with such a crossover. It should be also noted that the inducement of a metastable ferromagnetic or ferrimagnetic states was observed also in some rare-earth based intermetallics (see [38], for instance) and in molecular magnets containing high anisotropic 3d ions [45]. The Ising character of Fe ions in $Fe_{0.5}TiS_2$ is responsible not only for the presence of the field-induced metamagnetic AF-F phase transition but also for the persistence of the metastable ferromagnetic state after removal of the field and for the high coercivity (~ 50 kOe) at low temperatures. The giant magnetic coercivity ($H_c$ ~ 90-100 kOe) associated with non-zero orbital moment and Ising character of Fe ions was also observed in $LuFe_2O_4$ [46] having a layered crystal structure as well. According to [18], Fe ions in this compound exhibit an extremely high orbital moment (~ 1 $\mu_B$).

The results obtained in the present work show that the $Fe_{0.5}TiS_2$ compound exhibits an AF ground state despite the presence of additional ferromagnetic-like anomalies on the temperature dependences of the magnetization detected on the sample heat treated at 350°C. These anomalies as well as the partial hysteresis loops (shown in figure 5(c) and 7(d)) revealed for the heat-treated $Fe_{0.5}TiS_2$ sample by the isothermal measurements of the magnetization and magnetoresistance are suggested to originate in the appearance of short-range ferromagnetic correlations within the AF matrix with decreasing temperature below 75 K. We think that because of the observation of such kind anomalies, the $Fe_xTiS_2$ compounds with the Fe concentration around $x = 0.5$ were classified in some papers [16, 21, 22] as having a ferromagnetic order. The lower width of the partial hysteresis loops in the heat-treated $Fe_{0.5}TiS_2$ sample may be associated with a weaker exchange interaction within such ferromagnetic clusters in comparison with the exchange energy in the main AF ordered phase. It should be noted that the coexistence of a long-range antiferromagnetic order and spin-glass or cluster spin-glass was revealed in some diluted Ising systems, for instance, in $(Fe,Mg)Cl_2$ [46] and $PbFe_{0.5}Nb_{0.5}O_3$ [47] in which the Fe ions are substituted by non-magnetic Mg and Nb, respectively. In the case of $Fe_xTiS_2$ compounds the subsystem of Fe ions located between S-Ti-S tri-layers can be also considered as an Ising system diluted by vacancies.

In summary, we systematically studied the magnetization and resistivity behavior in the intercalated compound $Fe_{0.5}TiS_2$. The results manifest that $Fe_{0.5}TiS_2$ exhibits an antiferromagnetic ground state. Because of Ising-type spin states of Fe ions this compound undergoes the metamagnetic phase transition to the ferromagnetic state under application of the magnetic field below the Neel temperature. The field induced AF-F transformation of the magnetic structure leads to the substantial (~ 27%) reduction of the electrical resistivity. The inducement of the F state in $Fe_{0.5}TiS_2$ is found to be irreversible below 100 K. The magnetization reversal in the metastable F state is accompanied by the large hysteresis with the coercitive field up to 50 kOe at low temperatures. The magnetic structure of $Fe_{0.5}TiS_2$ is suggested to be of layered type with the strong ferromagnetic coupling within layers and much weaker AF exchange interactions between layers. The proposed magnetic structure is needed to be verified by detailed neutron diffraction studies. The results obtained show that the magnetic phase diagram of the $Fe_xTiS_2$ system should be revised especially in the region with high Fe concentrations.

**Acknowledgments**

This work was supported in part by the program of the Ural Branch of RAS (project № 12-T-1012).